\documentstyle[12pt]{article}
\def\fun#1#2{\lower3.6pt\vbox{\baselineskip0pt\lineskip.9pt
\ialign{$\mathsurround=0pt#1\hfil##\hfil$\crcr#2\crcr\sim\crcr}}}

\newcommand{\be}{\begin{equation}}
\newcommand{\ee}{\end{equation}}
\newcommand{\bd}{\begin{displaymath}}
\newcommand{\ed}{\end{displaymath}}
\newcommand{\ba}{\begin{array}}
\newcommand{\ea}{\end{array}}
\newcommand{\bt}{\begin{tabular}}
\newcommand{\et}{\end{tabular}}
\newcommand{\bc}{\begin{center}}
\newcommand{\ec}{\end{center}}

\begin{document}


\hfill {\large\bf US-FT/8--96}

\hfill {\bf hep-ph/9602348}

\vspace{0.5cm}

\begin{center}
{\huge\bf $D$,$D^*(2010)$ and $D^{*}_{2}(2460)$ charmed mesons production 
in hadron--hadron collisions}\\[0.5cm]
G.H.Arakelyan$^1$, C.Pajares and Yu.M.Shabelski$^2$\\[0.5cm]

{\it Departamento de Fizica de Particulas, Universidade de Santjago de 
Compostela,\\
15706-Santjago de Compostela,Spain}

\end{center}


\vspace{1.cm}

\begin{abstract}

A resently developed generalization of the  Quark--Gluon String Model to the 
case of bosonic resonances production in hadron--hadron collisions is used for 
the calculation of inclusive production of charmed mesons $D$, $D^*(2010)$ and 
$D^{*}_{2}(2460)$. A simple relation which determines the dependence of the 
charmed meson production cross sections on their spin J is obtained. It is 
shown that the theoretical predictions for the inclusive spectra and production
cross sections of these charmed mesons are confirmed experimentally with a 
reasonable accuracy.
\end{abstract}

\vfill

~~~~ {\large\bf US-FT/8--96}
~~~~  February 9, 1996

\noindent
\underline{\hspace{6.5cm}}

{\small $^1$ Theoretical Department, Yerevan Physics Institute,
Br.Alikhanyan's str. 2, Yerevan 375036, Armenia;
e--mail: argev@vxc.yerphi.am~~~~~argev@vxitep.itep.ru}

{\small $^2$ Petersburg Nuclear Physics Institute, Gatchina, St.Petersburg
188350, Russia; e--mail: shabel@vxdesy.desy.de}

\newpage

\section{Introduction}

~~~~~ Hadroproduction of the charmed particles is now being investigated in 
many experiments at different energies of colliding hadrons. Usually the heavy 
flavour production processes in hadron-hadron interactions at high energies 
are considered in the framework of perturbative QCD \cite{1,2}. However these 
processes can be also considered in the framework of phenomenological 
Quark--Gluon String Model (QGSM) \cite{3} -- \cite{7}, which is one of 
the nonperturbative approaches to the description of hadron production 
processes.

The QGSM considers the inelastic hadron-hadron collision as a two step process:
an interaction which breaks the hadron into coloured constituents followed by a 
fragmentation of these constituents via formation of strings or chains. The 
QGSM treatment of string formation is based on 1/N expansion of QCD in the 
framework of the Dual Topological Unitarization (DTU) scheme \cite{8}. In the 
leading order of 1/N expansion, the particles are produced in two chains or 
strings (one-pomeron exchange) which are linked between constituents of 
different hadrons. Each of this two chains has a planar topology. The 
contribution of several pomeron exchange is also significant.

In QGSM the properties of the fragmentation functions of a quark $G_q(z)$ and a 
diquark $G_{qq}(z)$ are determined directly from an analysis of the planar 
diagrams on the basis of reggeon diagram technique. This approach allows one to
express the behavior of the fragmentation functions at $z{\rightarrow} 0$ and 
$z{\rightarrow} 1$ in terms of intercepts of the known Regge trajectories. The 
comparison of the QGSM predictions with the experimental data shows that the 
model describes quite reasonably the inclusive production of the pseudoscalar 
mesons \cite{9} -- \cite{11}.
   
Recently the QGSM was generalized \cite{12} to the case of bosonic resonances 
production which are lying on the leading trajectories of the vector--tensor 
($V-T$) group ({\it $\rho$, $a_{2}$, $f$, $K^*$ .....}), and the 
functions of quark and diquark fragmentation $G^J_{q(qq)}(z)$ into bosonic 
resonances with arbitrary spin $J$ were considered. The functions 
$G^J_{q(qq)}(z)$ at $z{\rightarrow} 1$ can be expressed in terms of the 
residues of secondary Regge trajectories which correspond in the framework of 
DTU approach to the contribution of planar diagrams. Using the predictions 
of the model for the spin structure of planar diagrams \cite{13}, the relations
between the residues of leading trajectories of the  (V--T) group were 
obtained. According to these predictions, the interaction of a V--T group 
reggeon with hadrons has an universal form similar to the case of 
electromagnetic interaction. The hypotesis of the dominance of electromagnetic
type interaction in the planar part of hadronic amplitudes (reggeon-photon 
analogy \cite{12,13}) together with the predictions of dual Veneziano model 
\cite{14} for the reggeon--particles couplings allows one to fix the quantities
$G^J_{q(qq)}(z)$ at $z {\rightarrow} 1$. As a result, the simple relation 
which determines the dependence of the production cross sections on the spin 
$J$ of resonances can be obtained. Analysis given in \cite{12} shows that the 
model reproduces correctly the available data on the high energy Feynman-$x$ 
spectra of bosonic resonances with arbitrary spins lying on ${\rho}$-- and
$K^*$-- trajectories.

In the present paper we apply this approach to the cross sections of high spin 
charmed meson $D^*(2010)$ and $D^{*}_{2}(2460)$ production using the QGSM 
description of $D$ meson production cross section in $pp$ and $\pi p$ 
collisions \cite{4,6}. We describe the data on inclusive spectra and production
cross sections $D^*(2010) ((\frac {1}{2})^+1^- )$ -- meson in $\pi p$ 
\cite{15,16} and $pp$ \cite{17} collision. The predictions for 
$D^{*}_{2}(2460)$--meson production cross sections and inclusive spectra are 
presented.
 
\section{Model description}

~~~~~ 
The inclusive spectrum of a secondary hadron $h$ in the framework of the QGSM 
has the form \cite{6,9}:

\be
\label{1}
\frac{x_E}{\sigma_{in}}\frac{d\sigma^h}{dx}=\sum^{\infty}_{n=1}
w_n(s)\varphi^h_n(x)+V^{(1)}_D\varphi^{(1)}_D(x)+
V^{(2)}_D\varphi^{(2)}_D(x)~.
\ee
Here  $x_E=E/E_{max}$,
\be
\label{2}
w_n(s)=\sigma_n(s)/\sum^{\infty}_{n=1}\sigma_n(s)
\ee
is the probability of cut precisely $n$ pomerons, $\sigma_n(x)$ is the cross 
section of $n$--pomeron shower production and $\varphi^h_n(x)$ determines the 
contribution of a diagram with $n$ cut pomerons. Two last terms in (1) 
correspond to the diffraction dissociation contribution which is negligibly
small in the case of charmed meson production.

The expressions for $w_n(s)$ and corresponding parameter values
for $pp$ and $\pi p$ collisions are given in \cite{9}--\cite{11}.

 The function $\varphi^h_n(x)$ $(n>1)$ for $\pi p$ interaction can be
written in the form \cite{9}--\cite{11}:

\be
\label{3}
\varphi^h_n(x)=f^h_{\bar
q}(x_+,n)f^h_q(x_-,n)+f^h_q(x_+,n)f^h_{qq}(x_-,n)+2(n-1)f^h_s(x_+,n)
f^h_s(x_-,n)
\ee
and for baryon--proton interaction
\be
\label{4}
\varphi^h_n(x)=f^h_{qq}(x_+,n)f^h_q(x_-,n)+f^h_q(x_+,n)f^h_{qq}(x_-,n)+
2(n-1)f^h_s(x_+,n)f^h_s(x_-,n)~,
\ee
where $x_{\pm}=\frac{1}{2}[\sqrt{x^2+x^2_{\bot}}\pm x]$.

The functions $f^h_i(x,n)(i=q,q_s,\bar q,qq)$ in (\ref{3})--({4}) describe the 
contributions of the valence/sea quarks, antiquarks and diquarks, respectively.
They represent a convolution of quark/diquark momentum distribution functions 
$u_i(x,n)$ in the colliding hadrons with the fragmentation functions of 
quark/diquark into a secondary hadron $h$, $G^h_i(x,n)$:

\be
\label{5}
f_i(x,n)=\int^1_x u_i(x_1,n)G_i(x/x_1)dx_1~.
\ee

The projectile (target) contribution depends only on the valiable $x_+$
$(x_-)$.

The functions $f^h_i(x,n)(i=q,\bar q,qq,q_{sea})$ for $D$--mesons production 
in $\pi p$ and $pp$ collision together with full list of distribution 
$u_i(x,n)$s and fragmentation $G^h_i(x,n)$ functions are presented in \cite{6}.

Some general properties of fragmentation functions were discussed in \cite{12}. 
In the two limits, $z \rightarrow 0$ and  $z \rightarrow 1$, the behaviour of
the fragmentation functions $G^h_{q(qq)}(z)$ can be found from their Regge 
asymptotics:
     
\be
\label{6}
G^h_{q(qq)}(z)=a^h(1-z)^\gamma \;.
\ee
Here $\gamma$ is determined by the intercepts of correspondance Regge 
trajectories. For our purposes the most important is to consider the constant 
$a^h$, which is the value of function  $G^h_{q(qq)}(z)$ at  $z \rightarrow 0$ 
and does not depend on the type of initial quark $q$ (diquark $qq$). The 
constants $a^h$ are determined by the dynamics of the fragmentation of the 
string when $q\bar q$ pair is produced from the vacuum. For instanse, the 
SU(3)-flavour symmetry requires that $a^{\rho^+} = a^{\rho^-} = a^{\rho^0} =
a^{\rho}, ...$ and $a^{D^+} = a^{D^-} = a^{D^0} = a^{\bar D^0} = a^D$, ... A 
direct calculation of these constants can not be done in the framework of 
QGSM. In \cite{12} the ratios of the constants conserning light and strange 
mesons were estimated. As an approximation was assumed that the shapes of 
$x_F$-spectra of resonances produced in the quark-gluon string do not depend 
on their spins $J$. In the considered approximation, the functions of quark 
fragmentation into different resonances of $\rho$ and $K^*$ families can be 
expressed via functions of quark fragmentation into $\rho-$ or $K^*$-mesons 
\cite{12}:  

\be
\label{7}
G^J_{q(qq)}(z)=R_JG^{\rho,K^*}_{q(qq)}(z) \;,
\ee
where the quantity $R_J$ does not depend on the variable $z$. Analogously the 
relation for resonances of $D^*$ family with spin $J_c$ can be written in the 
similar form:

\be
\label{8}
G^{J_c}_{q(qq)}(z)=R_{J_c}G^{D^*}_{q(qq)}(z) \;.
\ee

The quantities $R_J$ can be expressed in terms of constants $a^J$:
$R_J=(a^J/a^V)^2$. Following to \cite{12} it is possible to express the 
quantities $a^J$ in terms of intersept $\alpha_{V}(0)$ of the trajectory
to which a resonance $J$ belongs:

\be
\label{9}
R_J\equiv(a^J/a^V)^2=\sigma^J/\sigma^V
   =\frac{(J+1)!}{(2J)!}(J-\alpha_V(0))^{J-1} \;.
\ee
 
According to the results \cite{12} for light and strange mesons we estimate
the connection between $D^*_2(2460) \equiv D^{**}$ and $D^*(2010) \equiv D^*$ 
residues:

\be
\label{10}
(a^{D^{**}})^2 \approx 0.84(a^{D^*})^2 \;.
\ee
 
Using the predictions of the resonance decay model \cite{18}, in \cite{12} the 
relations were obtained for the probabilities of the production of the light 
and strange pseudoscalar and vector mesons. In the case of the charmed 
mesons production such relation has the form

\be
\label{11}
(a^{D^*}/a^D)^2=\frac{<k^2_{\bot}>_{D}}{4m^2_q} \;,
\ee
where $m_q$ is the transverse mass of the constituent quark. The value 
$m_q=0.415\pm0.015$ GeV which was used in \cite{18} for light and strange mesons
leads for $D$-mesons to the wrong predictions, namely the multiplicity of 
produced $D$-mesons is predicted to be smaller than the multiplicity of 
$D^*(2010)$ that is impossible because all $D^*(2010)$ should decay into $D$. 
We found the value of parameter $m_q$ from the comparison with the data on $D$ 
and $D^*(2010)$ production cross sections and get a reasonable agreement with 
the value $m_q=0.65$ GeV. Taking into account the fact that 
$<{k^2_{\bot}}>_D=1$ GeV$^2$ and $a^D$=0.18 \cite{4,6} we found that

\be
\label{12}
a^{D^*} \approx 0.14
\ee
and using the relation (\ref{9})

\be
\label{13}
a^{D^{**}} \approx 0.12 \;.
\ee

\section{Comparison with experimental data and predictions of
the model}

~~~~~In this section we consider the description of the existing experimental 
data on $D$-- and $D^*$--mesons production in $\pi p$ and $pp$ collisions 
in the framework of the presented model. The predictions for 
$D^*_2(2460)$--meson inclusive spectra and production cross sections are also 
presented.

The experimental data on inclusive distributions of $D$--mesons were described 
in \cite{4} (see also \cite{3,5,6,7,8}), so we present only  comparsion with 
data \cite{19} on  $D$--mesons produced in the $pp$ --collision at 
$800\;GeV/c$. 

In Fig. 1 we present a comparison of the $x_F$--distribution for all 
$D$--mesons produced in $pp$ collision at 800 GeV/c \cite{19}. The QGSM 
result (solid curve) is in reasonable agreement with the data. The 
predictions for inclusive spectra of $D^{*}(2010)$ (dashed curve) and 
$D^{*}_2(2460)$ (dashed-dotted curve) resonances are also shown. 

The inclusive spectra of $D^*(2010)$--mesons  produced in $\pi^-p\to
D^{*+}/D^{*-}X$ and $\pi^-p\to D^{*0}/\bar D^{*0}X$ reactions at $360\;GeV/c$ 
\cite{16} are compared with our calculations in Fig. 2a, b respectively. It 
seems that the agreement with the data on  sum of spectra of $D^{*+}$ and 
$D^{*-}$ mesons (Fig. 2a, dashed curve) is rather good. As to the neutral 
mesons $D^{*0}$ and $\bar D^{*0}$ (Fig. 2b) the theoretical curve lies slightly
lower, but the experimental information is rather scarce. Fig. 2 contains also 
the theoretical curves for $D$-- and $D^*_2(2460)$--mesons spectra. The  
comparison of the QGSM calculation for $D$-- meson spectra with experimental 
data were given in \cite{4}.

Our predictions for inclusive spectra of different $D^*_2(2460)$--mesons 
produced in $pp$ colisions at 400 GeV/c are presented in Fig. 3.
  
The experimental data on the integral cross sections of $D$ and $D^*$ meson 
production in $pp$ and $\pi p$ collision  are compared with our calculations 
in the Table. We also give the prediction for $D^*_2(2460)$--meson cross 
section both in $pp$ and $\pi p$ collision.


The results given in the present paper show that the approach based on the QGSM
and reggeon--photon analogy developed in the \cite{12} may be used for the 
estimations of charmed meson production cross sections.

We are grateful to A.A.Grigoryan N.Ya.Ivanov and for stimulating discussions. 
Yu.M.Shabelski also thanks Direction General Politica Cientifica of Spain for 
financial support. This work was supported in part by the grant INTAS--93--0079.

\newpage

\bc
{\Large\bf Table }\\[0.3cm]

Experimental data and model calculations for $D$--, $D^*(2010)$-- and
$D^*_2(2460)$--meson production cross sections in $\pi p$ and $pp$ collisions.

\vspace{5mm}

\bt{|l|c|c|c|c|}\hline
~~~~~Reaction & Ref. & $P_L\;(GeV/c)$ & $\sigma_{exp}(\mu b)$ &
$\sigma_{theor}(\mu b)$\\ \hline

$pp\to D^+/D^-~X$ & \cite{19} & $800$ & $26\pm14$ & $23$ \\
&&&& \\

$pp\to D^0/\bar D^0~X$ & \cite{19} & $800$ & $22$\parbox{1cm}{$+4$\\
$-7$} & $31$ \\
&&&& \\

$pp\to D^{*+}/D^{*-}~X$ &         & $800$ &        & $13$  \\
&&&&\\

$pp\to D^{*0}/\bar D^{*0}~X$ &      & $800$ &      & $17$  \\
&&&&\\

$pp\to D^{**+}/D^{**-}~X$ &      & $800$ &    & $6.5$ \\
&&&&\\

$pp\to D^{**0}/\bar D^{**0}~X$ &  &  $800$ & $   $ & $9.0$ \\
&&&&\\

$pp\to D^{*+}/D^{*-}~X$ & \cite{17} & $400$ & $9.2\pm2.4$ & $6.9$ \\
&&&&\\

$pp\to D^{*0}/\bar D^{*0}~X$ & \cite{17} & $400$ & $5.8\pm2.7$ &
$10.0$ \\
&&&&\\

$pp\to D^{**+}/D^{**-}~X$ &     & $400$ &       & $3.5$ \\
&&&&\\

$pp\to D^{**0}/\bar D^{**0}~X$ &     & $400$ &      & $5.7$ \\
&&&&\\

$\pi^-p\to D^{*+}/D^{*-}~X$ & \cite{16} & $360$ &
$5.0$\parbox{1cm}{$+2.3$\\ $-1.8$} & $8.6$ \\ &&&&\\

$\pi^-p\to D^{*0}/\bar D^{*0}~X$ & \cite{16} & $360$ & $7.3\pm2.9$ &
$6.8$ \\
&&&&\\

$\pi^-p\to D^{**+}/\bar D^{**-}~X$ &    & $360$ & $   $ & $4.7$ \\
&&&&\\

$\pi^-p\to D^{**0}/\bar D^{**0}~X$ &    & $360$ & $    $ & $3.5$ \\
&&&&\\ \hline
\et
\ec

\newpage

\section*{Figure captions}

\noindent
\bt{p{1.5cm}p{14.5cm}}

Fig.1 & Inclusive distributions of all $D$--mesons in $pp$ interaction at 
$800\;GeV/c$ (solid curvee) and data \cite{19} together with the model 
predictions for $D^*(2010)$-- (dashed curve) and $D^*_2(2460)$-- mesons 
(dotted curve) distributions.\\[2mm]

Fig.2 & The $x_F$--dependence of $D^*$--meson production in $\pi^-p$ 
interactions at $360\;GeV/c$ (dashed curve) and data \cite{16}: a) 
$D^{*+}/D^{*-}$--mesons, b) $D^{*0}/\bar D^{*0}$--mesons. The curves for $D$-- 
(solid curve) and $D^*_2(2460)$--meson (dotted curve) distributions are also 
presented.\\[2mm]

Fig.3 & The prediction on $x_F$--dependence of inclusive spectra of 
different $D^*_2(2460)$--mesons in $pp$ collision at 400 GeV/c: solid curve 
-- $D^{*+,0}_2(2460)$, dashed curve -- $\bar D^{*0}_2(2460)$ and dashed-dotted 
curve -- $D^{*-}_2(2460)$.

\et


\newpage

\begin{thebibliography}{99}

\bibitem{1} P.Nason, S.Dawson and R.K.Ellis. Nucl. Phys. {\bf B303} (1988) 607.

\bibitem{2} G.Altarelli, M.Diemoz, G.Martin and P.Nason. Nucl.Phys. 
{\bf B308} (1988) 724.

\bibitem{3} A.B.Kaidalov and O.I.Piskounova. Yad. Fiz. {\bf43} (1986) 1545.

\bibitem{4} Yu.M.Shabelski. Yad. Fiz. {\bf55} (1992) 2227; L.Cifarelli, 
E.E\c{s}kut and Yu.M.Shabelski. Nuono Cimento. {\bf A106} (1993) 389.            

\bibitem{5} O.I.Piskounova. Yad. Fiz. {\bf56} (1993) 176; Phys. A. Nucl. 
{\bf56} (1993) 1094.

\bibitem{6} Yu.M.Shabelski. Surveys in High Energy Physics {\bf 9} (1995) 1.

\bibitem{7} G.H.Arakelyan and P.E.Volkovitski. Z. Phys. A {\bf 353} (1995) 87.

\bibitem{8} M Giafalloni, G.Marchezini and G.Venetsiano. Nucl. Phys. 
{\bf B98} (1975) 472.

\bibitem{9} A.B.Kaidalov and K.A.Ter--Martirosyan. Yad. Fiz. {\bf39}
(1984) 1545; ibid. {\bf40} (1984) 211.

\bibitem{10} A.B.Kaidalov and O.I.Piskounova. Yad. Fiz. {\bf41} (1985) 1278;
Sov. J. Nucl. Phys. {\bf41} (1985) 816.

\bibitem{11} Yu.M.Shabelski. Yad. Fiz. {\bf44} (1986) 186.

\bibitem{12} G.G.Arakelyan, A.A.Grigoryan, N.Ya.Ivanov and A.B.Kaidalov.
Z. Phys. C {\bf63} (1994) 137.

\bibitem{13} A.A.Grigoryan, N.Ya.Ivanov and A.B.Kaidalov. Yad. Fiz. {\bf36} 
(1982) 1490; Sov. J. Nucl. Phys. {\bf36} (1982) 867.

\bibitem{14} G.Veneziano. Nuono Cimento {\bf A57} (1968) 190.

\bibitem{15} S.Barlag et al. Z. Phys. C {\bf39} (1988) 451.

\bibitem{16} M.Aguilar--Benitez et al. Phys. Lett. {\bf169B} (1986) 106.

\bibitem{17} M.Aguilar--Benitez et al. Z. Phys. C {\bf40} (1988) 321.

\bibitem{18} A.A.Grigoryan, N.Ya.Ivanov. Yad. Fiz. {\bf43} (1986) 693;
Sov. J. Nuc. Phys. {\bf43} (1986) 442.

\bibitem{19} R.Ammar et al. Phys. Rev. Lett. {\bf61} (1988) 2185.

\end{thebibliography}
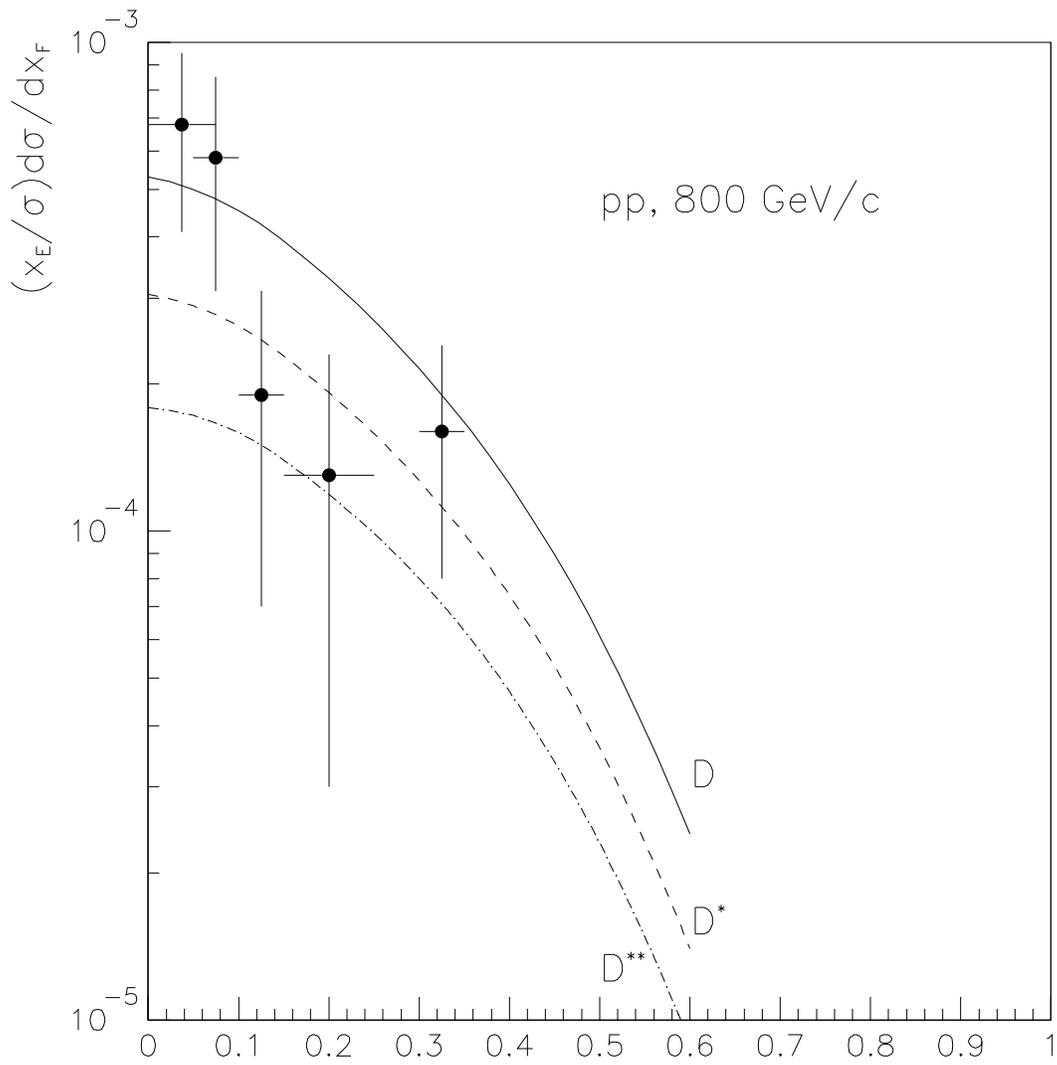
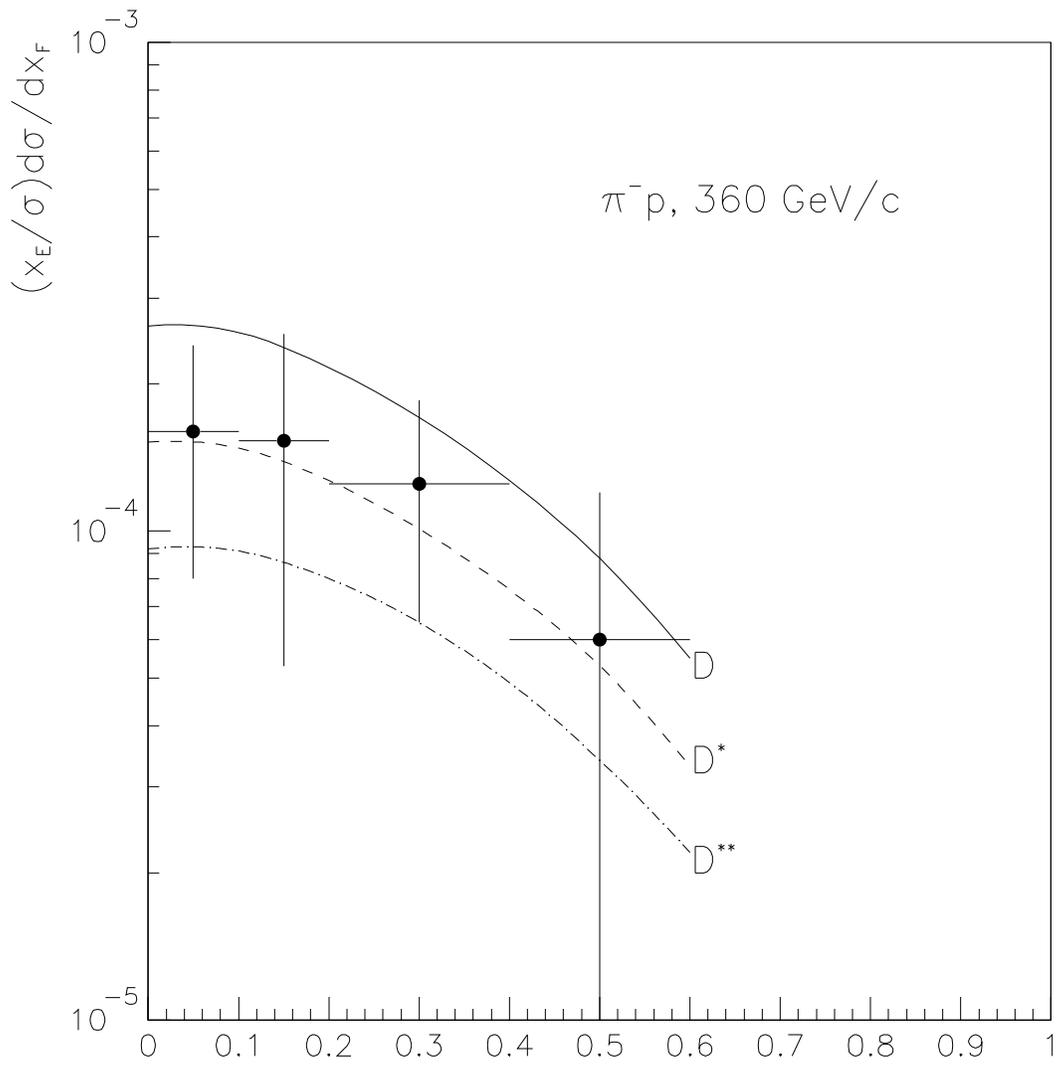
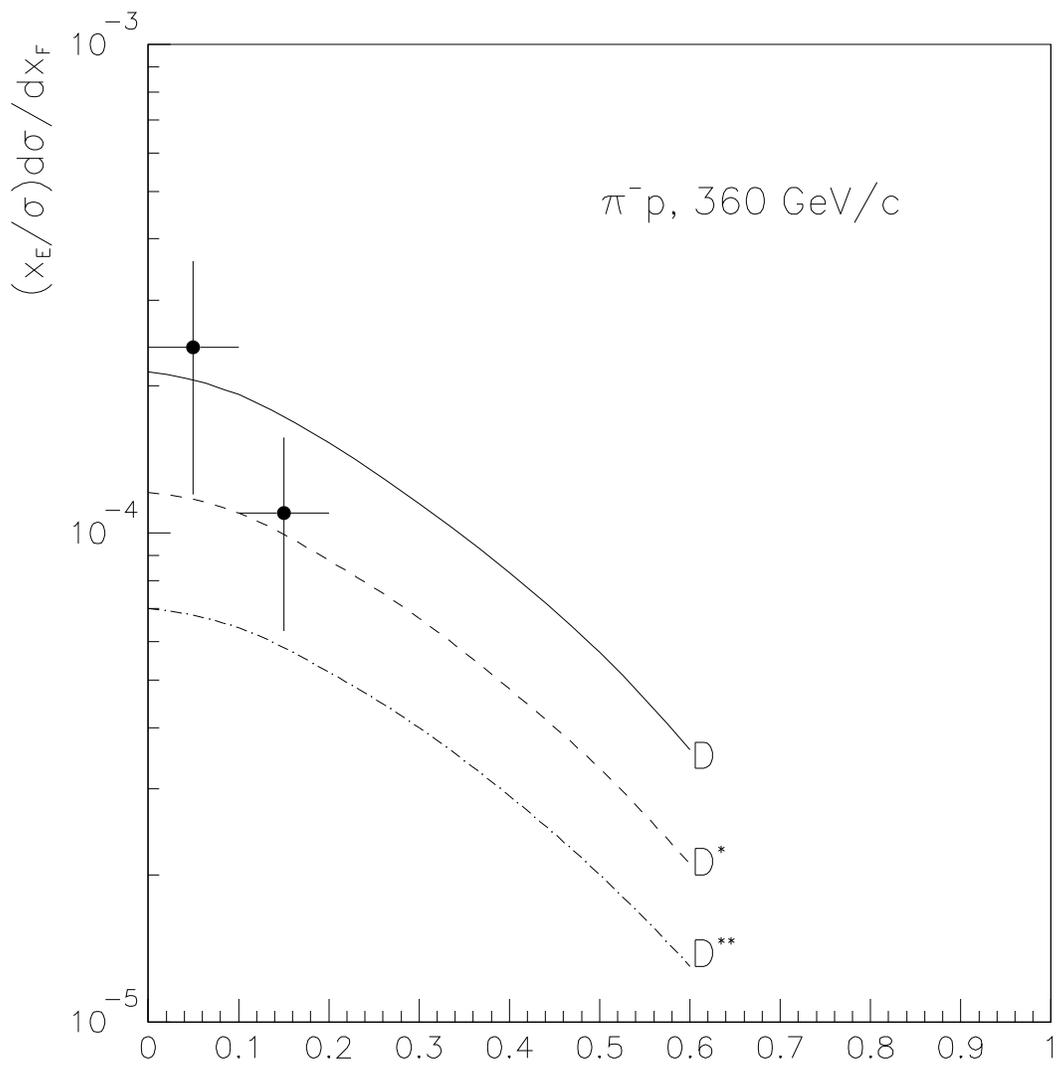
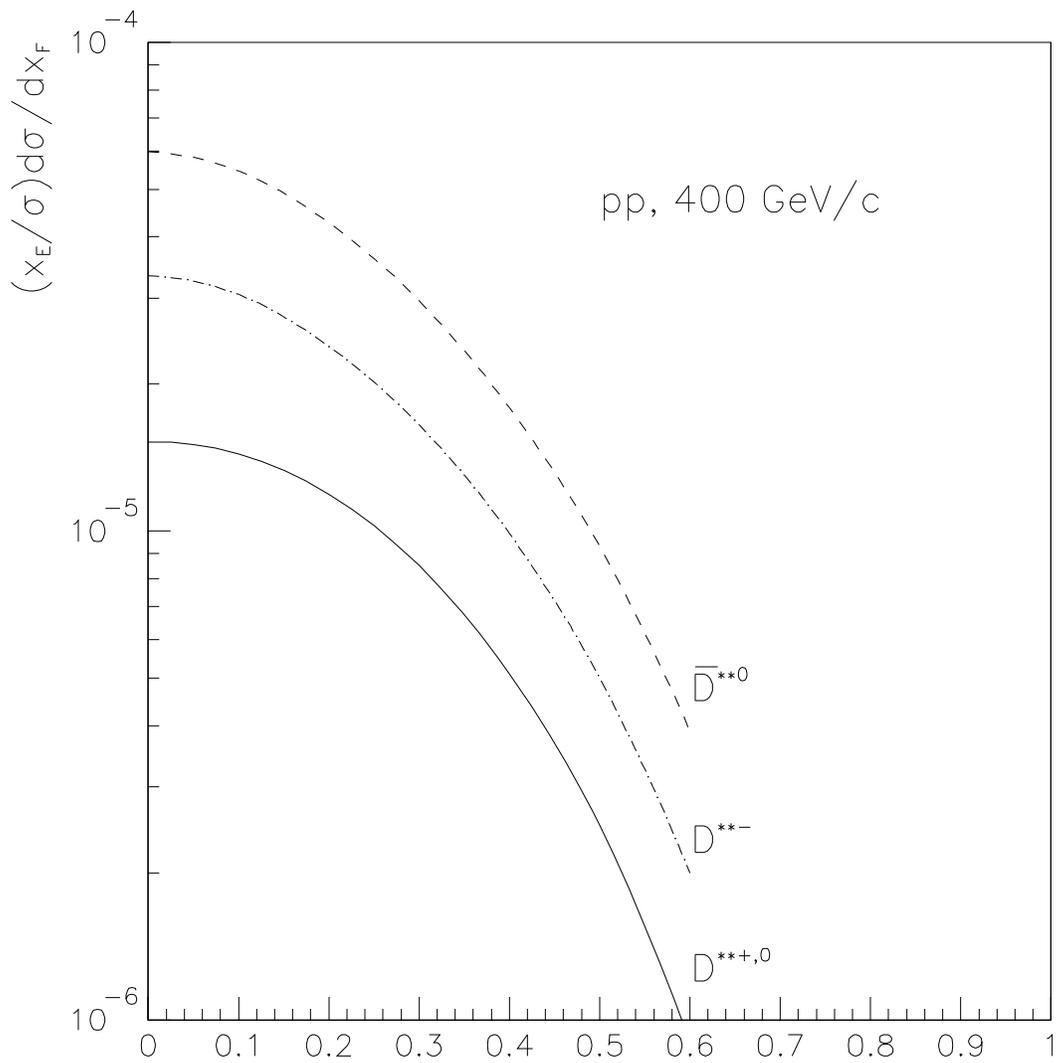
\end{document}